# Purely Electric-Field Control of Topological Magnetism in Two-Dimensional Magnets

Zhonglin He, Naafis Ahnaf Shahed, Kai Huang, Himanshu Mavani and Evgeny Y. Tsymbal*

*Department of Physics and Astronomy & Nebraska Center for Materials and Nanoscience, University of Nebraska, Lincoln, Nebraska 68588-0299, USA*

*Corresponding author: tsymbal@unl.edu

## Abstract

Electrical control of topological magnetism is central to realizing energy-efficient topological spintronics. Yet most electric-field approaches modify the competing magnetic interactions through a nonselective rearrangement of low-energy electronic states, yielding coarse magnetic phase control that often requires external magnetic fields to stabilize topological quasiparticles, while few schemes solely based on electric fields are restricted to specific conducting materials. Here, we establish a general approach for purely electric-field control of topological magnetism, in which an applied electric field electrostatically dopes a selected orbital-angular-momentum-polarized band edge of a two-dimensional (2D) van der Waals (vdW) magnetic semiconductor via proximity to an adjacent nonmagnetic vdW metal. We show that the resulting electrostatic doping predominantly tunes magnetic anisotropy of the 2D magnet, while leaving exchange interaction and Dzyaloshinskii-Moriya interaction nearly unchanged, thereby reversibly driving the system through ferromagnetic, skyrmion, spiral, and bimeron phases. We demonstrate this mechanism for $CrBr_3$/graphene and $Cr_2Ge_2Te_6$/$TaS_2$ vdW heterostructures hosting electron and hole pockets of different orbital characters. These results establish a broadly applicable strategy for purely electric-field control in topological spintronics.

## Introduction

Topological magnetism, rooted in the real-space winding of noncollinear spin configurations, has emerged as a fertile setting for exploring topological spintronics [1-4]. Protected by nontrivial topology, these magnetic quasiparticles exhibit appealing characteristics as next-generation information carriers, including robustness against continuous deformation, nanometric size, and low threshold currents [5-8]. Translating these features into practical spintronic devices, however, hinges on achieving effective control over these spin textures, in particular their reversible creation and annihilation as well as transitions between distinct topological phases [9-15]. To this end, various external stimuli, such as spin-transfer/orbit torques [10,11], magnetic/electric fields [12-15], and laser excitation [16], have been harnessed to manipulate topological magnetism. Among these, electric-field control stands out as a distinct route to energy-efficient spintronic functionality without driving currents [12-15]. Identifying effective strategies for electrically tunable topological magnetism is therefore of central importance to bridging fundamental topological physics and future spintronic applications.

Two-dimensional (2D) magnetic materials have opened new frontiers in the investigation of topological magnetism [17-21]. The emergence of topological magnetism in 2D lattices stems from a delicate balance of competing magnetic interactions [17]. Confining these interactions to the reduced-dimensional limit [21], together with gate-tunable interfacial electrostatics [22,23], makes 2D magnets a novel platform for electric-field control of topological magnetism. Recent studies have demonstrated electric-field control of transitions between skyrmion and bimeron phases, as well as the creation and annihilation of topological spin textures in thin-film and 2D magnetic systems [14,24-30]. Yet existing approaches have generally modified the competing magnetic interactions through a nonselective rearrangement of low-energy electronic states. The states brought near the Fermi level are dictated by the material-specific band structure rather than selected according to their contributions to the magnetic interactions. The resulting control over magnetic phases is correspondingly coarse. In most cases, the electric field only shifts the balance among the magnetic interactions into a range favorable for topological spin textures, while an external magnetic field remains necessary to stabilize isolated topological quasiparticles [13,24-27,30]. Moreover, the few purely electric-field proposals remain tied to specific material systems [15,28,31]. A general strategy for purely electric-field control of topological magnetism has yet to be established.

In this work, we establish a general approach for purely electric-field control of topological magnetism in 2D materials. This approach is based on electrostatic doping of a selected band edge in a heterostructure formed by a 2D van der Waals (vdW) magnetic semiconductor and an adjacent nonmagnetic vdW metal. Guided by a minimal $k \cdot p$ model, we show that weak electrostatic doping of a band edge carrying finite orbital angular momentum $\langle \mathbf{L} \rangle$, characterized by out-of-plane (OP) $\langle \mathbf{L}_z \rangle$ or in-plane (IP) $\langle \mathbf{L}_x \rangle$ polarization, substantially modifies the magnetic anisotropy while leaving the exchange interaction and Dzyaloshinskii-Moriya interaction (DMI) nearly unchanged. Such selective tunability enables reversible creation and annihilation of topological spin textures as well as switching between skyrmion and bimeron phases without an external magnetic field. Using first-principles calculations and atomistic spin-model simulations, this approach is further demonstrated in two distinct vdW heterostructures, $CrBr_3$/graphene and $Cr_2Ge_2Te_6/TaS_2$. In $CrBr_3$/graphene, interfacial charge transfer from graphene partially fills the conduction-band edge near the K point, forming an $\langle \mathbf{L}_x \rangle$-polarized electron pocket. An applied electric field tunes the occupation of this pocket and reverses the magnetic anisotropy from OP to IP, driving a reversible phase evolution through OP-ferromagnetic (FM), skyrmion, trivial spiral, bimeron, and IP-FM phases. In $Cr_2Ge_2Te_6/TaS_2$, interlayer charge transfer shifts the $\langle \mathbf{L}_z \rangle$-polarized valence-band edge near the Γ point to the Fermi level, forming a hole pocket. Electric-field-induced changes in the occupation of this pocket similarly reverse the magnetic anisotropy and drive an analogous sequence of magnetic phases.

## Results

Topological magnetism arises from competing magnetic interactions [17,25], typically including exchange interaction $J_{ij}$, magnetic anisotropy $K$ and DMI $d_{ij}$, as captured by the following spin Hamiltonian,

$$H_{\text{spin}} = -\sum_{i,j} J_{ij} \mathbf{S}_i \cdot \mathbf{S}_j - \sum_{i,j} d_{ij} \hat{\mathbf{d}} \cdot (\mathbf{S}_i \times \mathbf{S}_j) - K \sum_i (\mathbf{S}_i \cdot \hat{\mathbf{z}})^2. \quad (1)$$

Here, $\mathbf{S}_i$ is the spin vector at site $i$. $\hat{\mathbf{z}}$ is the unit vector perpendicular to the plane of the 2D magnetic lattice, and $\hat{\mathbf{d}}$ denotes the unit vector along the DMI vector. Electric-field control of topological magnetism is expected to occur through changes in these magnetic parameters. In 2D materials, electrostatic gating intrinsically provides such an approach by modulating the occupation of electronic states near the Fermi level [31-33]. This mechanism is particularly effective for magnetic anisotropy $K$, which is intimately tied to the low-energy electronic states and thus susceptible to carrier modulation [34]. For a band edge carrying a finite orbital angular momentum $\langle \mathbf{L} \rangle$, characterized by an OP component $\langle \mathbf{L}_z \rangle$ or IP component $\langle \mathbf{L}_x \rangle$, weak electrostatic doping can induce a sizable change in $K$ [35-37]. The central requirement is therefore not carrier modulation alone, but placing an $\langle \mathbf{L} \rangle$-polarized band edge within a field-tunable range around the Fermi level. In a vdW heterostructure composed of a 2D magnetic semiconductor and a nonmagnetic vdW conducting layer, weak interlayer coupling can preserve much of the orbital character of the semiconductor band edge [38]. Interlayer charge transfer shifts the selected band edge to the Fermi level, where partial filling or depletion forms an electron or hole pocket [38-40]. An OP electric field then changes the relative electrostatic potential of the two layers and hence tunes the pocket occupation [41]. However, how such weak electrostatic doping reorganizes the competing magnetic interactions and thereby transforms the associated topological spin textures, remains an open question. To address this question, we employ a minimal $k{\cdot}p$ model for an $\langle \mathbf{L}_z \rangle$- or $\langle \mathbf{L}_x \rangle$-polarized low-energy band edge whose Hamiltonian is given by:

$$H = \xi C(k_x^2 + k_y^2) - \Delta_{\mathrm{ex}}\, \mathbf{S} \cdot \boldsymbol{\sigma} - \boldsymbol{\Omega}_{\mathrm{SOC}} \cdot \boldsymbol{\sigma} + \alpha_R (k_y \sigma_x - k_x \sigma_y). \quad (2)$$

Here, the first term describes a parabolic band edge, with $\xi = +1$ ($-1$) corresponding to a conduction (valence) band. $\Delta_{ex}$ is half the spin splitting that couples the carriers with the local magnetization $\mathbf{S}(\mathbf{r})$ [see **Fig. 1(a)**]. The $\langle \mathbf{L} \rangle$ polarization sets an effective onsite spin-orbit coupling (SOC) field $\boldsymbol{\Omega}_{\mathrm{SOC}} = \lambda_{\mathrm{eff}} \langle \mathbf{L} \rangle / 2$, oriented along $z$ or $x$ for $\langle \mathbf{L}_z \rangle$ or $\langle \mathbf{L}_x \rangle$, respectively. $\lambda_{\mathrm{eff}}$ describes the strength of SOC and denotes the spin splitting induced by SOC, as shown in **Fig. 1(a)**. The last term represents Rashba SOC, which generates DMI.

In a weak electrostatic doping regime, the rigid band approximation is appropriate: electrostatic doping shifts the Fermi level relative to the band edge. As a representative case, we consider an $\langle \mathbf{L}_x \rangle$-polarized conduction band forming an electron pocket. Parameters for the low-energy Hamiltonian are included in **Supplementary Table 2**. The electron-doping-induced magnetic anisotropy $\Delta K$ is obtained from the energy difference of the doped electrons between OP-FM and IP-FM phases (see **Supplementary Note 1**). As shown in **Fig. 1(a)**, at the same electron doping, the IP-FM and OP-FM Fermi levels differ by an offset $\Delta\mu$. This offset reflects lower energy of IP-FM, thereby favoring IP magnetization and giving rise to a negative $\Delta K$. For exchange interaction and DMI, a spin spiral $\mathbf{S}(\mathbf{r}) = (\sin qx, 0, \cos qx)$ is introduced to extract the doping induced $\Delta J$ and $\Delta d$ [see **Supplementary Note 1**]. **Fig. 1(b)** shows the evolution of $\Delta K$, $\Delta J$ and $\Delta d$ as functions of electron doping. Clearly, weak doping produces a sizable change in $\Delta K$, whereas the corresponding $\Delta J$ and $\Delta d$ remain comparatively small.

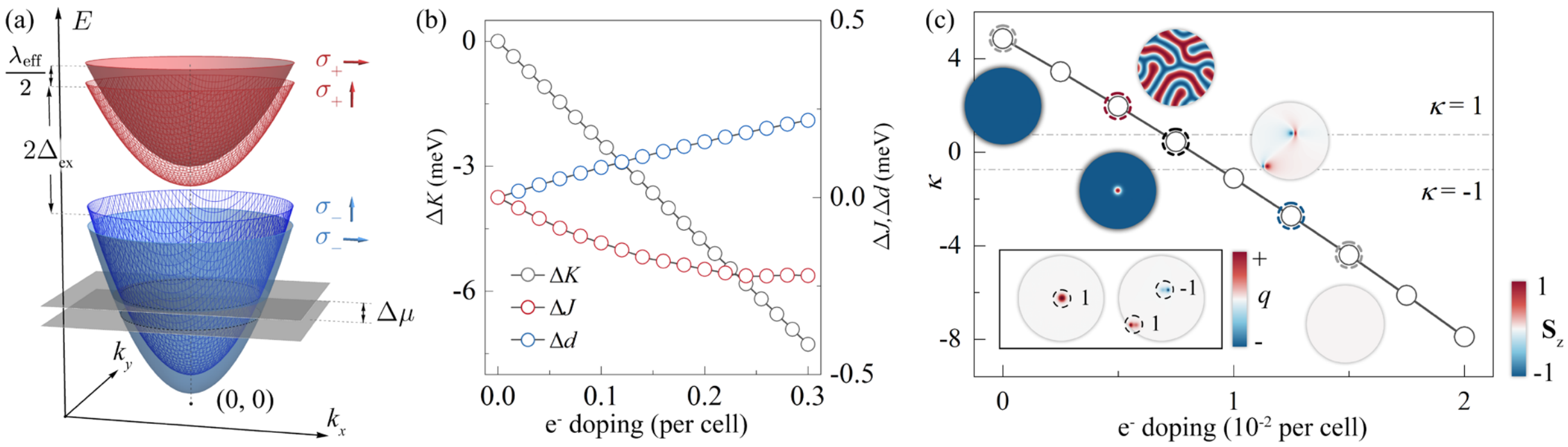

**Fig. 1**. (a) Schematic of the $\langle \mathbf{L}_x \rangle$ polarized electron pocket. ($\sigma_+$, red) and ($\sigma_-$, blue) denote the two spin channels. Arrows denote IP-FM and OP-FM bands. The gray planes indicate the IP-FM and OP-FM Fermi levels differ by an offset $\Delta\mu$ at the same electron doping. (b) Magnetic anisotropy $\Delta K$, exchange interaction $\Delta J$ and DMI $\Delta d$ as functions of electron doping, represented by black, red, and blue circles, respectively. (c) Parameter $\kappa$ as a function of electron doping. Topological phase boundaries $\kappa = 1$ and $\kappa = -1$ are marked by dashed lines. Evolution of spin textures as a function of electron doping is also depicted. OP spin component is denoted by color. Inset shows the topological charge density $q(\mathbf{r})$ of skyrmion and bimeron phases with the integer topological charges ($Q = \pm 1$) of topological quasiparticles labeled. The color bar indicates the topological charge density.

To determine how doping influences the spin textures, we superimpose $\Delta K$, $\Delta J$ and $\Delta d$ onto an undoped spin Hamiltonian of hexagonal lattice shown in **Supplementary Eq. (S6)** with a sizable DMI ($J_{NN}$ = 5 meV, $d_{NN}$ = 0.5 meV and $K$ = 0.2 meV). To quantify the resulting phase evolution, we introduce a dimensionless parameter, $\kappa=\left(\frac{4}{\pi}\right)^2\frac{2J_{NN}K}{3d_{NN}^2}$, which controls the emergence of topological magnetism. In general, $|\kappa| < 1$ favors spin spirals, $|\kappa| > 1$ allows topological magnetism to emerge, and $|\kappa| >> 1$ stabilizes a FM state [28]. As shown in **Fig. 1(c),** increasing the electron doping drives $\kappa$ across these phase boundaries inducing topological phase transitions.

These predictions are further confirmed by atomistic spin-model simulations. With the increase in electron doping, the spin texture evolves through a rich sequence of phase transitions, including OP-FM, skyrmion, trivial spiral, bimeron, and IP-FM phases [see **Fig. 1(c)**]. Topological charge $Q=\frac{1}{4\pi}\iint \mathbf{S}(\mathbf{r})\cdot\left[\frac{\partial \mathbf{S}(\mathbf{r})}{\partial x}\times\frac{\partial \mathbf{S}(\mathbf{r})}{\partial y}\right]d^2\mathbf{r}$ is calculated using the discrete formula in **Eq. (3)** to quantify topological properties of these textures. As illustrated in the inset of **Fig. 1(c)**, for skyrmion and bimeron phases, the topological charges $Q$ are integers and topological charge density $q(\mathbf{r})=\frac{1}{4\pi}\mathbf{S}(\mathbf{r})\cdot\left[\frac{\partial \mathbf{S}(\mathbf{r})}{\partial x}\times\frac{\partial \mathbf{S}(\mathbf{r})}{\partial y}\right]$ peaks around topological quasiparticles, confirming their topological nature. These results demonstrate that the dominant carrier-induced change in magnetic anisotropy alters the balance among the magnetic interactions and drives this sequence of phase transitions without an external magnetic field.

To design or search for 2D heterostructures capable of realizing this mechanism, two key ingredients are essential. The first is a band edge carrying finite $\langle \mathbf{L}\rangle$, whose weak doping produces a sizable change in magnetic anisotropy. The second is an adjacent weakly coupled nonmagnetic vdW metal that shifts the selected band edge to the Fermi level and enables electrostatic control of its occupation. A sufficiently large DMI, arising from inversion symmetry breaking either at the interface or within the magnetic semiconductor itself, is additionally required to stabilize topological spin textures. In practice, one should therefore first identify a 2D magnetic semiconductor with such a conduction- or valence-band edge and then select an adjacent nonmagnetic vdW metal that provides the required charge transfer and electrostatic tunability.

Now we consider realistic materials that satisfy these requirements. Monolayer $CrBr_3$, an OP-FM semiconductor, is a natural candidate. As shown in **Supplementary Fig. 1(a)**, the valence band of $CrBr_3$ exhibits a sizable OP orbital angular momentum $\langle \mathbf{L}_z\rangle$, with a maximum near the Γ point. In contrast, the conduction band exhibits a weaker orbital polarization with a strong momentum dependence. **Fig. 2(c)** shows the magnitude $\left|\langle \mathbf{L} \rangle\right|$ and the ratio $\gamma = \dfrac{\langle \mathbf{L}_x \rangle}{\langle \mathbf{L}_z \rangle}$ of the conduction band throughout the Brillouin zone. Although $\langle \mathbf{L} \rangle$ in the conduction band is predominantly $\langle \mathbf{L}_z \rangle$-polarized over most of the Brillouin zone, it becomes primarily $\langle \mathbf{L}_x \rangle$ near the K point.

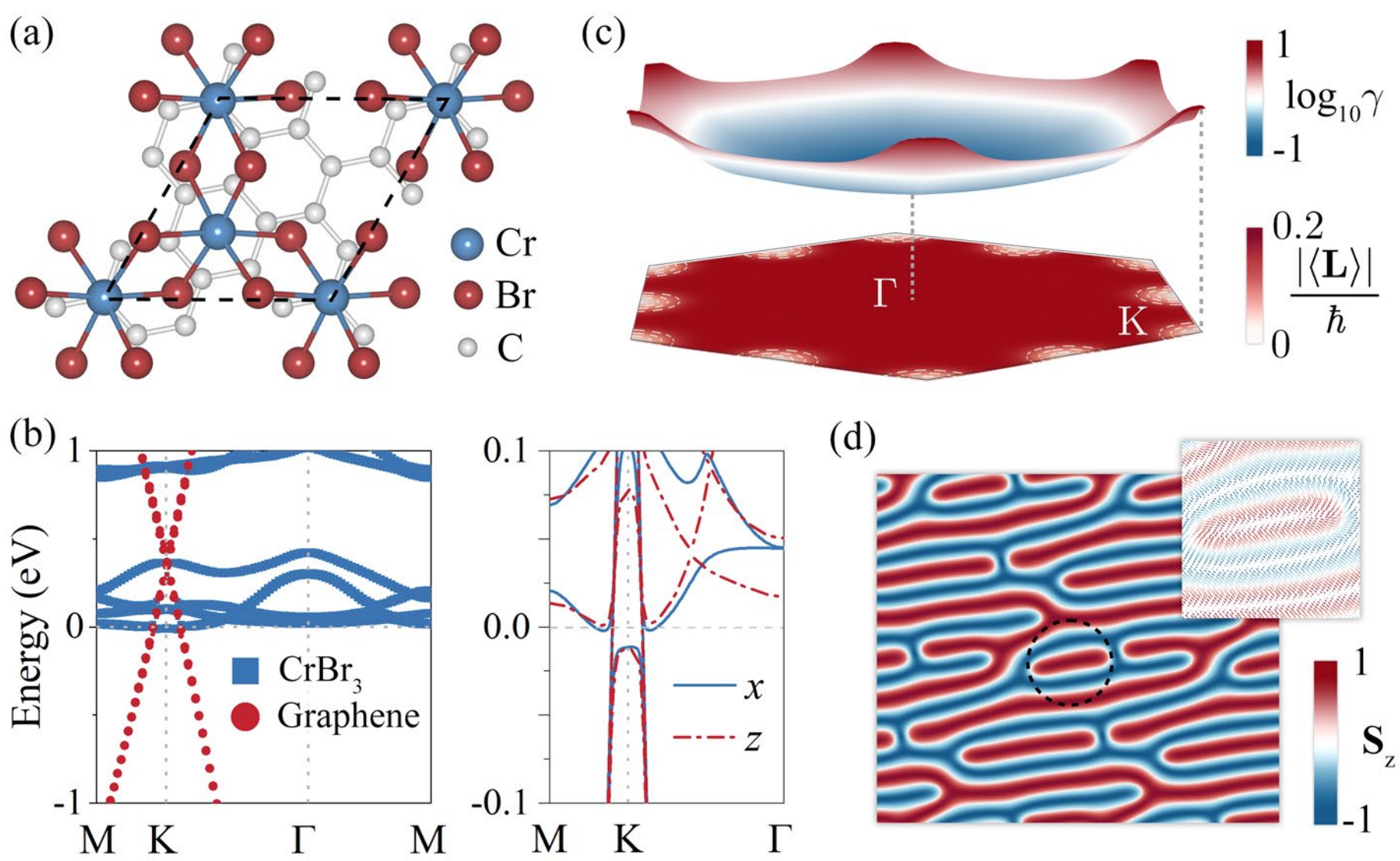

**Fig. 2**. (a) Crystal structure of $CrBr_3$/graphene vdW heterostructure. The unit cell is denoted by the dashed line. Cr, Br and C atoms are shown as blue, red, and gray spheres. (b) Left: layer resolved band structure of $CrBr_3$/graphene vdW heterostructure, with the weight on $CrBr_3$ and graphene indicated by the blue and red markers, respectively. Right: band structures of the IP-FM (solid blue line) and OP-FM (dashed red line) $CrBr_3$/graphene vdW heterostructure near the K point. (c) Distribution of $\gamma$ and $\left|\langle \mathbf{L} \rangle\right|$ for the conduction band of monolayer $CrBr_3$ across the Brillouin zone, with the color bars denoting their values. (d) Spin texture of $CrBr_3$/graphene vdW heterostructure. The inset is an enlarged view of spin textures marked by the dashed circle. The color bar indicates the OP spin component and arrows denote the IP spin component.

To place the band edge near the Fermi level and control its occupation, we interface $CrBr_3$ with monolayer graphene. We find that a $\sqrt{7} \times \sqrt{7}$ graphene supercell accommodates the $CrBr_3$ unit cell with a minimal lattice mismatch of 1.2%. To identify the stable stacking configuration of the $CrBr_3$/graphene, we calculate the relative energies of different stacking configurations by interlayer sliding [see **Supplementary Fig. 2(a)**]. The ground state stacking configuration, shown in **Fig. 2(a)**, is then used in all subsequent calculations. **Fig. 2(b)** shows the electronic band structure of the IP-FM $CrBr_3$/graphene vdW heterostructure. A small amount of electron transfer from graphene to $CrBr_3$ shifts the conduction-band edge near the K point to the Fermi level, forming an $\langle \mathbf{L}_x \rangle$ -polarized electron pocket. To investigate the magnetic properties of the heterostructure, we employ the spin Hamiltonian in **Supplementary Eq. (S11)** including nearest-neighbor (NN) and next-nearest-neighbor (NNN) couplings. By mapping the total energies of multiple magnetic configurations onto this spin Hamiltonian (see **Supplementary Note 3**), we extract magnetic anisotropy, exchange interaction and DMI, as detailed in **Supplementary Table 3**. We find that magnetic anisotropy favors IP magnetization. For the exchange interactions, both $J_{NN}$ and $J_{NNN}$ are positive, indicating FM coupling. A finite DMI arises from the broken inversion symmetry at the heterostructure interface.

Next, we perform atomistic spin-model simulations to investigate the magnetic textures of the heterostructure. For a zero applied electric field ($E$ = 0), we find that the competition between the magnetic parameters stabilizes a labyrinth domain phase, as shown in **Fig. 2(d).** It is seen, however, that the spirals are on the verge of a topological phase transition as they start to break up and tend to form topological quasiparticles, as indicated by the inset of **Fig. 2(d)**.

To explore topological phase transitions in the $CrBr_3$/graphene heterostructure, we first examine how the electric field alters the magnetic parameters defined in **Supplementary Eq. (S11)**. We note that an OP electric field acts primarily by modulating the occupation of the selected electron pocket rather than through structural reconstruction. Even at the largest fields considered (±0.1 V/Å), structural relaxation produces only a small change in the interlayer spacing. We therefore fix the atomic structure of the $CrBr_3$/graphene vdW heterostructure for subsequent calculations.

**Fig. 3(a)** shows the evolutions of magnetic parameters of $CrBr_3$/graphene as functions of electric field normalized to their zero-field values, $J_0$, $d_0$ and $K_0$. A positive electric field, $E > 0$, is defined as pointing from $CrBr_3$ to graphene. As the electric field increases from −0.05 to 0.1 V/Å, $J$ and $d$ fluctuate within a narrow range, while $K$ varies significantly. For $E$ between −0.05 and −0.01 V/Å, $K > 0$, favoring OP magnetization. With increasing electric field, $K$ decreases and approaches zero near $E = -0.01$ V/Å. Upon further increasing the electric field, $K$ changes sign, favoring IP magnetization, and its magnitude continues to increase up to 0.1 V/Å.

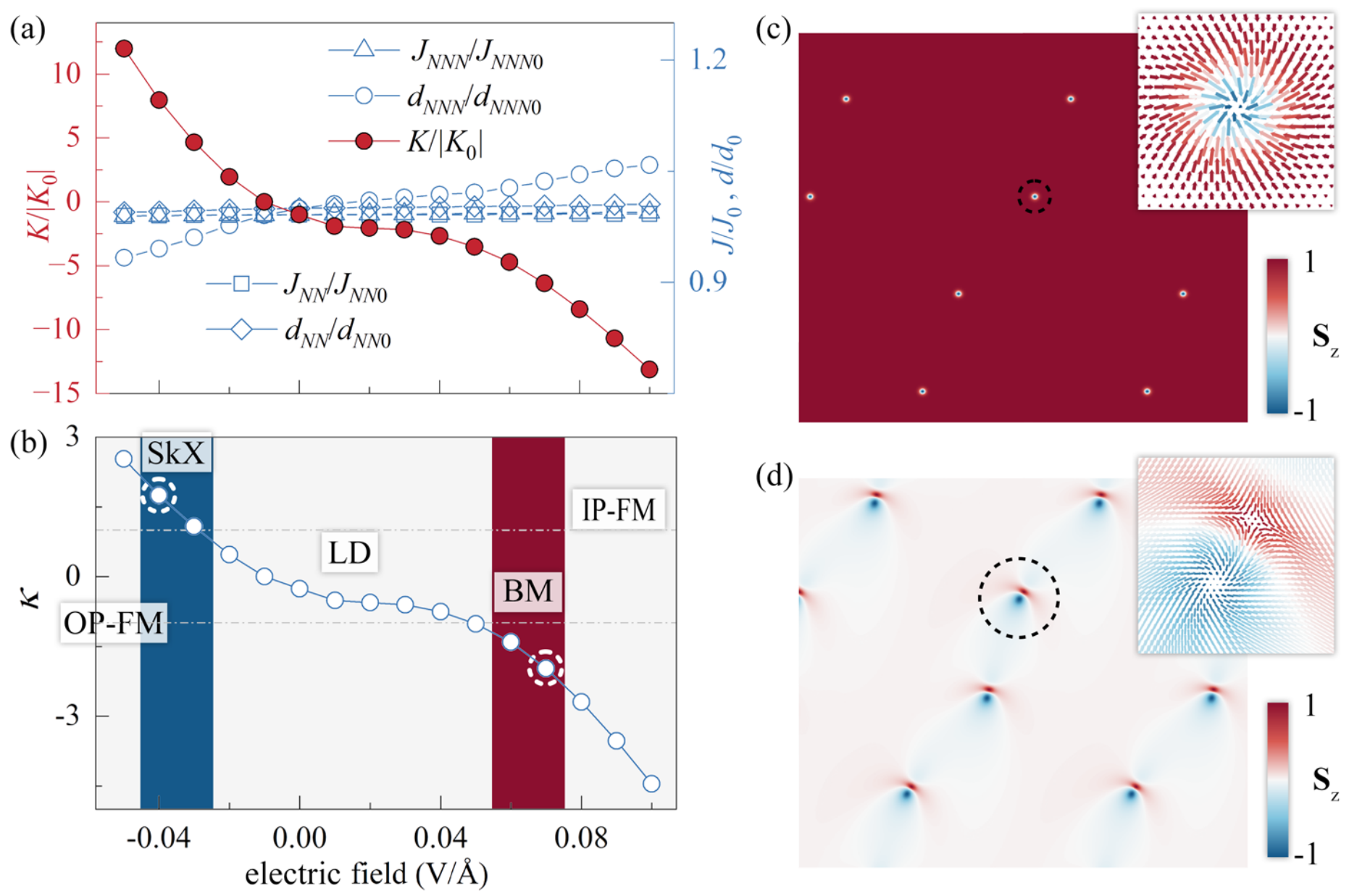

**Fig. 3**. (a) The normalized magnetic anisotropy $K/|K_0|$ (red filled circles), the normalized NN and NNN exchange interactions ($J_{NN}/J_{NN0}$, hollow squares; $J_{NNN}/J_{NNN0}$, hollow triangles) and DMI ($d_{NN}/d_{NN0}$, hollow diamonds; $d_{NNN}/d_{NNN0}$, hollow circles) as functions of an applied electric field. (b) The parameter $\kappa$ as a function of electric field. Topological phase boundaries $\kappa = 1$ and $\kappa = -1$ are marked by dashed lines. Skyrmion, labyrinth domain and bimeron phases are denoted by SkX, LD and BM, respectively. The two dashed circles mark the electric fields, at which spin textures are shown in (c) and (d). (c, d) Spin textures at electric fields of (c) −0.04 and (d) 0.07 V/Å. Insets are enlarged views of topological quasiparticles marked by the dashed circle. The color bar indicates the OP spin component and arrows denote the IP spin component.

The electric tunability of $K$ in $CrBr_3$/graphene is driven by the field-dependent occupation of the electron pocket. Accordingly, for $E < 0$, the electric field depletes the $\langle \mathbf{L}_x \rangle$-polarized electron pocket of $CrBr_3$, thereby enhancing OP magnetic anisotropy. By contrast, for $E > 0$, the electric field fills this pocket, favoring IP magnetic anisotropy, in agreement with the minimal model. The effect of electric field on the electron pocket is further confirmed by the calculated band structure, shown in **Supplementary Fig. 3**. This strong tunability of the magnetic anisotropy, combined with the nearly invariant $J$ and $d$, establishes the $CrBr_3$/graphene vdW heterostructure as a promising platform for realizing purely electric-field control of topological magnetism.

To quantify the phase transitions, we evaluate the electric-field dependence of the dimensionless parameter $\kappa$, defined in **Supplementary Eq. (S9)**, using the obtained magnetic parameters [see **Fig. 3(b)**]. As noted above, $|\kappa| = 1$ marks the boundary for the emergence of topological magnetism. For the electric field ranging from −0.05 to −0.03 V/Å, $\kappa$ remains above the topological phase boundary at $\kappa = 1$. As the electric field is increased to −0.02 V/Å, $\kappa$ crosses the boundary, indicating a transition from the skyrmion phase to a trivial spiral phase. With further increasing electric field, $\kappa$ continues to decrease and changes sign near $E = -0.01$ V/Å. At $E = 0.05$ V/Å, $\kappa$ crosses the topological phase boundary at $\kappa = -1$ and reenters the topological regime, marking a transition from the trivial spiral to the bimeron phase. When the electric field exceeds $E = 0.07$ V/Å, $\kappa$ decreases further, driving the system out of the topological regime into the trivial IP-FM phase.

These predicted electric-field-driven phase transitions are further supported by atomistic spin-model simulations. As shown in **Supplementary Fig. 4(a)**, an OP-FM state is observed at $E = -0.05$ V/Å. As the electric field increases, isolated skyrmions emerge within the FM background, forming a skyrmion phase [see **Fig. 3(c)**]. Upon increasing the electric field to −0.02 V/Å, these skyrmions pair and condense into topologically trivial spin spirals, which remain stable within the range from −0.02 to 0.04 V/Å. With further increase of the electric field along $+z$, the spirals break up into meron-antimeron loops and eventually evolve into a bimeron phase over the range from 0.06 to 0.07 V/Å [see **Fig. 3(d)**]. For electric fields exceeding 0.07 V/Å, the bimerons annihilate into the background, leaving an IP-FM state [see **Supplementary Fig. 4(b)**].

**Supplementary Figs. 5(a, b)** show the topological charge $Q$ as well as topological charge density $q(\mathbf{r})$ of these topological phases. The integer-quantized $Q$, together with the correspondence between localized $q(\mathbf{r})$ and the topological quasiparticles, confirms their topological nature. Moreover, the reversibility of the phase transition is demonstrated in **Supplementary Figs. 6(a, b)**. Upon removal of the electric field, the topological spin texture transforms into a labyrinth domain state, whereas reapplying the field regenerates the topological quasiparticles.

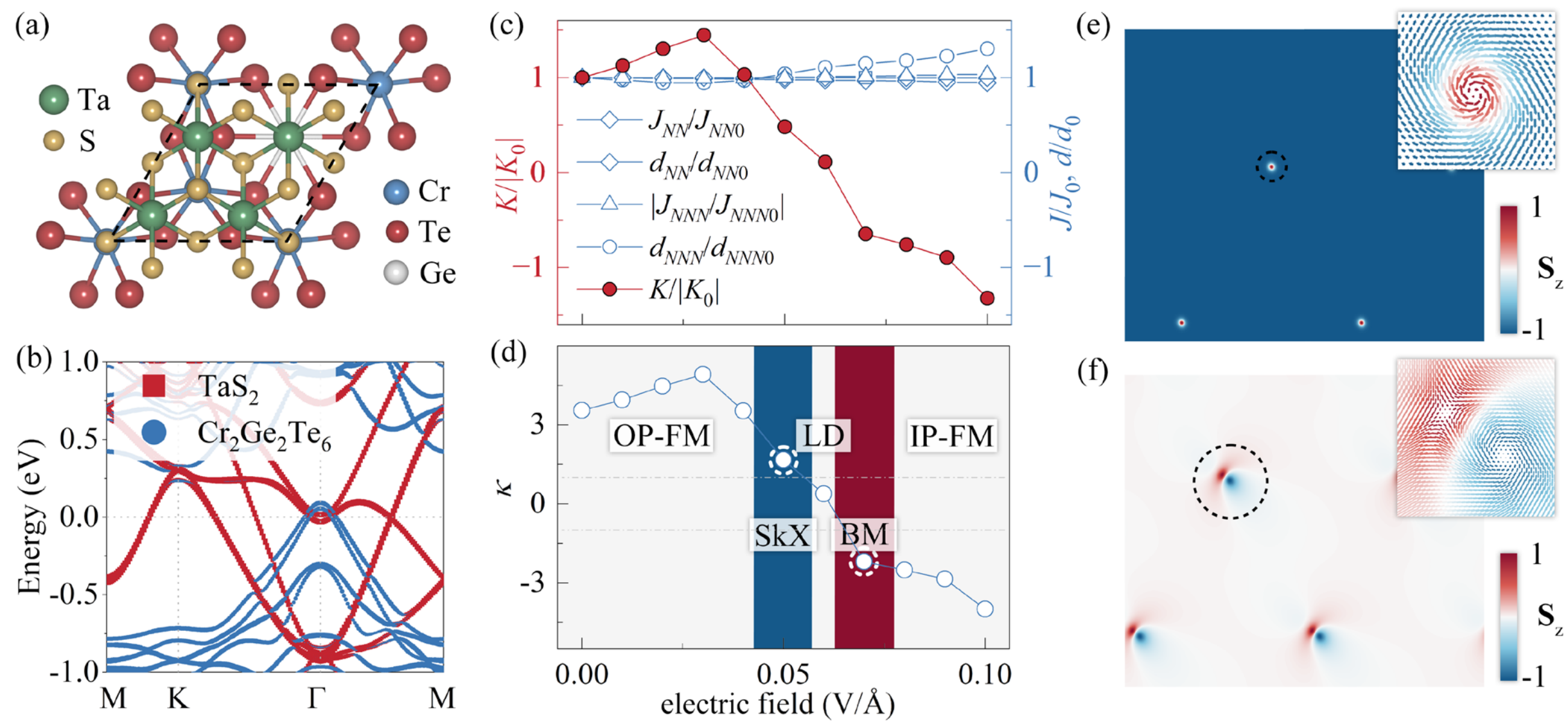

**Fig. 4**. (a) Crystal structure of $Cr_2Ge_2Te_6/TaS_2$ vdW heterostructure. The unit cell is denoted by the dashed line. Cr, Te, Ge, Ta and S atoms are shown as blue, red, gray, green and yellow spheres. (b) Layer-resolved band structure of $Cr_2Ge_2Te_6/TaS_2$, with the weight on $Cr_2Ge_2Te_6$ and $TaS_2$ indicated by the blue and red markers, respectively. (c) The normalized magnetic anisotropy $K/|K_0|$ (red filled circles), the normalized NN and NNN exchange interactions ($J_{NN}/J_{NN0}$, hollow squares; $|J_{NNN}/J_{NNN0}|$, hollow triangles) and DMI ($d_{NN}/d_{NN0}$, hollow diamonds; $d_{NNN}/d_{NNN0}$, hollow circles) as functions of the electric field. (d) Parameter $\kappa$ as a function of electric field. Topological phase boundaries $\kappa = 1$ and $\kappa = -1$ are marked by dashed lines. Skyrmion, labyrinth domain and bimeron phases are denoted by SkX, LD and BM, respectively. The two dashed circles mark the electric fields, at which spin textures are shown in (e) and (f). (e, f) Spin textures at electric fields of (e) 0.05 and (f) 0.07 V/Å. Insets are enlarged views of topological quasiparticles marked by the dashed circle. The color bar indicates the OP spin component and arrows denote the IP spin component.

To establish the generality of this approach, we apply it to a distinct magnetic semiconductor, monolayer $Cr_2Ge_2Te_6$, which exhibits an OP-FM ground state with an $\langle \mathbf{L}_z \rangle$-polarized valence band [see **Supplementary Fig. 1(b)**]. Monolayer $TaS_2$ is selected as the adjacent nonmagnetic vdW metal based on the lattice matching. A 2 × 2 $TaS_2$ supercell matches the unit cell of $Cr_2Ge_2Te_6$ with a lattice mismatch of only 2%. The ground-state stacking configuration is identified through interlayer sliding [see **Supplementary Fig. 2(b)**]. In contrast to $CrBr_3$/graphene, interlayer charge transfer shifts the valence-band edge of $Cr_2Ge_2Te_6$ at the Γ point to the Fermi level, forming an $\langle \mathbf{L}_z \rangle$-polarized hole pocket, as shown in **Fig. 4(b)**.

We then investigate the magnetic properties of the heterostructure. The magnetic parameters extracted from **Supplementary Eq. (S11)** are summarized in **Supplementary Table 3**. The magnetic anisotropy favors OP magnetization, consistent with prior modeling of monolayer $Cr_2Ge_2Te_6$ [42]. The positive $J_{NN}$ and substantially weaker negative $J_{NNN}$ favor the FM state. Similar to $CrBr_3$/graphene, a finite DMI emerges due to broken inversion symmetry at the interface. Using the obtained magnetic parameters, we perform atomistic spin-model simulations to determine the resulting spin textures of $Cr_2Ge_2Te_6/TaS_2$. As shown in **Supplementary Fig. 4(c)**, the ground state is OP-FM.

Next, we examine the influence of an electric field on the magnetic parameters. Similar to $CrBr_3$/graphene, the applied electric field acts primarily by modulating the occupation of the selected hole pocket, while structural relaxation produces only a small change in the interlayer spacing even at the largest field considered (0.1 V/Å). We therefore use a fixed geometry in the subsequent calculations. A positive electric field, $E > 0$, is defined as pointing from $TaS_2$ to $Cr_2Ge_2Te_6$. As $E$ increases from 0 to 0.1 V/Å, $J$ and $d$ vary only weakly, whereas $K$ changes substantially. As shown in **Fig. 4(c)**, $K$ changes slowly up to $E = 0.03$ V/Å, then decreases rapidly and changes sign by $E = 0.07$ V/Å. Upon further increasing the field, $K$ continues to decrease.

This behavior can be understood from the electronic structure of the hole pocket. Its exchange splitting is nearly zero because the states forming the pocket originate predominantly from the ligand Te atoms. In the absence of an applied electric field, the hole pocket in $Cr_2Ge_2Te_6/TaS_2$ is nearly fully occupied [see **Supplementary Fig. 7(b)**]. Consequently, the initial effect of the electric field on $K$ is weak, consistent with the minimal model, as discussed in **Supplementary Note 2**. For $E > 0$, the electric field progressively depletes electrons from the $\langle \mathbf{L}_z \rangle$-polarized hole pocket of $Cr_2Ge_2Te_6$ and shifts the corresponding pocket of the IP-FM state downward in energy [see **Supplementary Figs. 7(c, d)**]. This relative shift increasingly favors the IP-FM state, thereby driving the pronounced reduction and eventual sign reversal of the magnetic anisotropy. Thus, the strong electric-field tunability of the magnetic anisotropy, together with the nearly invariant $J$ and $d$, establishes the $Cr_2Ge_2Te_6/TaS_2$ vdW heterostructure as another promising platform for purely electric-field control of topological magnetism.

To characterize the phase transitions, we examine the evolution of $\kappa$ as a function of electric field. As shown in **Fig. 4(d)**, $\kappa$ initially increases slightly as the field increases from 0 to 0.03 V/Å and then decreases towards the topological phase boundary at $\kappa = 1$. At $E = 0.06$ V/Å, $\kappa$ crosses this boundary, indicating a transition from the skyrmion phase to the trivial spiral phase. With a further increase in the electric field, $\kappa$ continues to decrease and changes sign. At $E = 0.07$ V/Å, $\kappa$ crosses the second topological phase boundary at $\kappa = -1$ and reenters the topological regime, indicating a transition from the trivial spiral to the bimeron phase. Upon further increasing the electric field, the system eventually leaves the topological regime and transitions to the trivial IP-FM phase.

To verify these predicted phase transitions, we perform atomistic spin-model simulations. As the electric field increases, isolated skyrmions emerge from the zero-field FM state, forming a skyrmion phase [see **Fig. 4(e)**]. With further increasing electric field, these skyrmions pair and condense into trivial spin spirals at $E = 0.06$ V/Å. At $E = 0.07$ V/Å, these spirals break up into meron-antimeron pairs and condense into a bimeron phase [see **Fig. 4(f)**]. For $E > 0.07$ V/Å, the bimerons annihilate, leaving an IP-FM phase [see **Supplementary Fig. 4(d)**]. The topological charge $Q$ and charge density $q(\mathbf{r})$ shown in **Supplementary Figs. 5(c, d)** further confirm the topologically nontrivial character of these topological phases. The reversibility of these electric-field driven phase transitions is demonstrated in **Supplementary Figs. 6(c, d)**, where switching the electric field off and on drives the spin textures from the topological state to the OP-FM state and back.

## Discussion

The proposed approach places electrical control of topological magnetism in a broader class of carrier-mediated magnetoelectric effects, but it differs in an important respect from previously explored approaches. Electric fields have been used to modify exchange interaction, DMI, or magnetic anisotropy through electrostatic doping [34,35,43], interfacial hybridization [28,36], ferroelectric polarization [15,25-27,31], and strain [13,14]. In most proposed or demonstrated systems, the nonselective rearrangement of low-energy electronic states causes several magnetic interactions to change simultaneously, and an external magnetic field is still needed to select or stabilize topological magnetism. Here, orbital-angular-momentum polarization provides a physical criterion for selecting a band edge that makes a sizable contribution to magnetic anisotropy:

changing the occupation of the resulting pocket produces a large variation of magnetic anisotropy while exchange interaction and DMI remain comparatively robust. The resulting change in the balance among these interactions connects OP-FM, skyrmion, trivial spiral, bimeron, and IP-FM states without an assisting magnetic field.

This viewpoint also suggests a materials-design strategy extending beyond the two examples considered here. The essential ingredients are a 2D magnetic semiconductor with a conduction- or valence-band edge carrying a sizable and directionally polarized orbital angular momentum, a sufficiently large DMI either generated by interfacial inversion symmetry breaking or intrinsic to the magnetic semiconductor, and an adjacent weakly coupled nonmagnetic vdW metal that transfers a small amount of charge and positions the selected band edge close to the Fermi level. The metal is not merely a passive electrode: its electronic structure and band alignment determine charge transfer and the gate-voltage range for filling or depleting the pocket. These criteria can be screened using the orbital character of the band edge in the isolated magnet and the band alignment of candidate heterostructures. They should be applicable to other chromium halides and tellurides, Janus magnets, and magnetic semiconductors interfaced with graphene, metallic dichalcogenides, or other vdW conductors with low density of states.

A realistic implementation could use an hBN-encapsulated dual-gate device. For $CrBr_3$/graphene, graphene can serve as both the adjacent conducting layer and local electrode, while an hBN-separated graphite gate controls the carrier density and OP displacement field. Fully encapsulated few-layer $CrBr_3$ devices [44] and graphene/$CrBr_3$ vertical stacks [45] have already been fabricated, providing a practical starting point for the proposed geometry. An analogous stack can be assembled for $Cr_2Ge_2Te_6$/$TaS_2$, with $TaS_2$ contacted as the metallic layer and an hBN-separated graphite gate on the opposite side. The calculated field scale, up to about 0.1 V/Å (1 V/nm), corresponds to displacement fields routinely targeted in thin-hBN-gated vdW devices, although the actual gate voltage will depend on dielectric thickness, quantum capacitance, and electrostatic screening. Dual gating is particularly valuable because it can separate a carrier-density effect from a pure displacement-field effect. The strong carrier-induced change in magnetic anisotropy observed in $Cr_2Ge_2Te_6$ under heavy electrolyte gating [43] supports this control principle, while weak doping emphasized here should avoid exchange reconstruction.

The predicted phase evolution can be tested by combining magnetic imaging with independent measurements of magnetic anisotropy. Polar Kerr microscopy or magnetic circular dichroism can track the OP magnetization, while cryogenic scanning NV magnetometry, already demonstrated for $CrBr_3$ [44], can resolve gate-dependent domain and skyrmion-like textures. Bimerons, with their IP background and weaker stray fields, may require vector-field reconstruction or complementary vector-sensitive imaging. Ferromagnetic resonance can determine gate-dependent magnetic anisotropy, with magnetotransport providing additional evidence for transitions between OP and IP magnetization. Observation of reversible transitions between OP-FM, skyrmion, trivial spiral, bimeron, and IP-FM phases together with an anisotropy sign change would directly test the proposed mechanism.

## Summary

In summary, our results establish electric-field-tunable occupation of orbital-angular-momentum-polarized electron or hole pockets as a general mechanism for controlling topological magnetism in 2D magnetic semiconductors adjacent to a nonmagnetic vdW metal. The resulting strong modulation of magnetic anisotropy, with exchange interaction and DMI remaining comparatively robust, drives the system across successive topological phase boundaries, enabling reversible transitions among OP-FM, skyrmion, trivial spiral, bimeron, and IP-FM phases without an external magnetic field. First-principles calculations combined with atomistic spin-model simulations demonstrate this mechanism for two distinct vdW heterostructures, $CrBr_3$/graphene and $Cr_2Ge_2Te_6/TaS_2$, involving electron and hole pockets of different orbital character. These results establish a broadly applicable route to the creation, annihilation, and reversible switching of topological spin textures using electric fields alone, providing a foundation for energy-efficient, all-electrical topological spintronics.

## Methods

**First-principles calculations.** Our first-principles calculations are performed based on density functional theory (DFT) using the projector augmented-wave method [46], as implemented in the Vienna Ab initio Simulation Package (VASP) [47,48]. For the exchange-correlation interaction, the generalized gradient approximation (GGA) in the form of Perdew-Burke-Ernzerhof functional is utilized [49]. The cutoff energy is set to 500 eV, and the convergence criterion of total energy is set to $10^{-7}$ eV. To avoid interaction between adjacent layers, a vacuum space of 30 Å is adopted. Structures are fully relaxed until the force on each atom is less than $10^{-2}$ eV/Å, and a $7 \times 7 \times 1$ *k*- point mesh is used to sample the Brillouin zone, except for the magnetic parameter calculations. The obtained lattice parameters are illustrated in **Supplementary Table 1**. To reduce the influence of strain on the magnetic properties, the lattice parameters of the magnetic sublayer are fixed when constructing the heterostructure, while the metallic sublayers are adjusted to match them. The zero-damping DFT-D3 method of Grimme is utilized for treating the interaction in vdW heterostructures [50], together with a dipole correction.

To precisely capture the magnetic properties of $CrBr_3$ and $Cr_2Ge_2Te_6$, the GGA+U method is employed to describe the strong correlated correction of Cr-3*d* electrons [51]. Following previous works, the value of U is chosen to be 0.5 eV for the Cr-3*d* electrons in $Cr_2Ge_2Te_6$ [52] and 1.0 eV for those in $CrBr_3$ [42]. Within magnetic parameters calculations, noncollinear DFT calculations including SOC are performed with magnetization orientation constrained, using an $N_1 \times N_2 \times 1$ k-point mesh with $N$ chosen such that $N_{1,2} \times a \approx 90$ Å, where $a$ is the corresponding lattice parameter.

**Atomistic spin-model simulations.** The atomistic spin-model simulations are performed using the VAMPIRE package [53] based on the spin Hamiltonian of **Supplementary Eqs. (S6)** and **(S11)**, and atomistic Monte Carlo simulations. A Monte Carlo approach is used to implement field-cooling protocols for determining the magnetic ground states and for modeling spin dynamics under applied electric fields. All field-cooling simulations are performed by gradually cooling down the system from its initial disordered state at high temperature ($T \gg T_c$) to 0 K with a step of 1 K. We obtain stable spin textures using a 300 × 300 supercell with periodic conditions. The iteration step is set to $10^7$.

The atomistic spin-model simulations are performed on a rhombic sample defined by $0 \leq x \leq L_0$ and $0 \leq y \leq \frac{\sqrt{3}L_0}{2}$, where $L_0$ is the side length of the rhombic sample. However, for visualization convenience, we map the rhombic sample onto a rectangular frame by applying a lattice transformation, with $\boldsymbol{a}'_1 = \boldsymbol{a}_1 + \boldsymbol{a}_2$ and $\boldsymbol{a}'_2 = \boldsymbol{a}_1 - \boldsymbol{a}_2$.

The topological charge $Q$ in 2D lattices is described as

$$Q = \frac{1}{4\pi}\sum_n q_s^n, \tag{3}$$

where $q_s^n$ is topological number density and has the form of $\tan\frac{q_s^n}{2} = \frac{\mathbf{S}_i^n \cdot (\mathbf{S}_j^n \times \mathbf{S}_k^n)}{1+\mathbf{S}_i^n \cdot \mathbf{S}_j^n + \mathbf{S}_j^n \cdot \mathbf{S}_k^n + \mathbf{S}_k^n \cdot \mathbf{S}_i^n}$.

Since the honeycomb lattice consists of two triangular sublattices, and the spin texture considered here is ferromagnetic and spatially smooth, the topological charge is evaluated on one triangular sublattice. $n$ runs over all equilateral triangles formed by three NNN Cr-Cr pairs across the triangular sublattice. $\mathbf{S}_i^n$, $\mathbf{S}_j^n$ and $\mathbf{S}_k^n$ are the three spin vectors of the n-th equilateral triangle in the anticlockwise lattice [54].

## Conflict of Interests

The authors declare no competing financial interest.

## Acknowledgements

This work was supported by the U.S. Department of Energy, Office of Science, Basic Energy Sciences under Award No. DE-SC0026103 (development of a minimal *k·p* model and density-functional calculations), the National Science Foundation through the EPSCoR RII Track-1 program (NSF Grant No. OIA-2044049) (evaluation of magnetic parameters and spin-model simulations), and the UNL Grand Challenges catalyst award "Quantum Approaches Addressing Global Threats" (computing a topological charge). Computations were performed at the University of Nebraska Holland Computing Center.

## References

1. Woo, S. Elusive spin textures discovered. *Nature* **564**, 43–44 (2018).

2. Göbel, B., Mertig, I. & Tretiakov, O. A. Beyond skyrmions: review and perspectives of alternative magnetic quasiparticles. *Phys. Rep.* **895**, 1 (2021).

3. Nagaosa, N. & Tokura, Y. Topological properties and dynamics of magnetic skyrmions. *Nat. Nanotechnol.* **8**, 899 (2013).

4. Wiesendanger, R. Nanoscale magnetic skyrmions in metallic films and multilayers: a new twist for spintronics. *Nat. Rev. Mater.* **1**, 16044 (2016).

5. Bogdanov, A. N. & Panagopoulos, C. Physical foundations and basic properties of magnetic skyrmions. *Nat. Rev. Phys.* **2**, 492 (2020).

6. Caretta, L., Mann, M., Büttner, F., Ueda, K., Pfau, B. et al. Fast current-driven domain walls and small skyrmions in a compensated ferrimagnet. *Nat. Nanotechnol.* **13**, 1154 (2018).

7. Zang, J., Mostovoy, M., Han, J. H. & Nagaosa, N. Dynamics of skyrmion crystals in metallic thin films. *Phys. Rev. Lett.* **107**, 136804 (2011).

8. Kanazawa, N., Seki, S. & Tokura, Y. Noncentrosymmetric magnets hosting magnetic skyrmions. *Adv. Mater.* **29**, 1603227 (2017).

9. Seki, S., Yu, X. Z., Ishiwata, S. & Tokura, Y. Observation of skyrmions in a multiferroic material. *Science* **336**, 198 (2012).

10. Jonietz, F. et al. Spin transfer torques in MnSi at ultralow current densities. *Science* **330**, 1648–1651 (2010).

11. Büttner, F. et al. Field-free deterministic ultrafast creation of magnetic skyrmions by spin-orbit torques. *Nat. Nanotechnol.* **12**, 1040–1044 (2017).

12. Sun, W. et al. Manipulation of magnetic skyrmion in a 2D van der Waals heterostructure via both electric and magnetic fields. *Adv. Funct. Mater.* **31**, 2104452 (2021).

13. Ba, Y. et al. Electric-field control of skyrmions in multiferroic heterostructure via magnetoelectric coupling. *Nat. Commun.* **12**, 322 (2021).

14. Wang, Y. et al. Electric-field-driven non-volatile multi-state switching of individual skyrmions in a multiferroic heterostructure. *Nat. Commun.* **11**, 3577 (2020).

15. Huang, K., Shao, D.-F. & Tsymbal, E. Y. Ferroelectric control of magnetic skyrmions in two-dimensional van der Waals heterostructures. *Nano Lett.* **22**, 3349–3355 (2022).

16. Finazzi, M. et al. Laser-induced magnetic nanostructures with tunable topological properties. *Phys. Rev. Lett.* **110**, 177205 (2013).

17. Lv, X. et al. Distinct skyrmion phases at room temperature in two-dimensional ferromagnet $Fe_3GaTe_2$. *Nat. Commun.* **15**, 3278 (2024).

18. Liang, J., Wang, W., Du, H., Hallal, A., Garcia, K. et al. Very large Dzyaloshinskii-Moriya interaction in two-dimensional Janus manganese dichalcogenides and its application to realize skyrmion states. *Phys. Rev. B* **101**, 184401 (2020).

19. Ding, B., Li, Z., Xu, G., Li, H., Hou, Z. et al. Observation of magnetic skyrmion bubbles in a van der Waals ferromagnet $Fe_3GeTe_2$. *Nano Lett.* **20**, 868–873 (2020).

20. Cao, T., Shao, D.-F., Huang, K., Gurung, G. & Tsymbal, E. Y. Switchable anomalous Hall effects in polar-stacked 2D antiferromagnet $MnBi_2Te_4$. *Nano Lett.* **23**, 3781–3787 (2023).

21. Han, M.-G., Garlow, J. A., Liu, Y., Zhang, H., Li, J. et al. Topological magnetic-spin textures in two-dimensional van der Waals $Cr_2Ge_2Te_6$. *Nano Lett.* **19**, 7859–7865 (2019).

22. Huang, C., Han, J., Wang, J., Jiang, J., Qu, Z. et al. Electric-field switching of interlayer magnetic order in a van der Waals heterobilayer via spin-electric potential. *Nat. Commun.* **16**, 10379 (2025).

23. Yao, F., Liao, M., Gibertini, M., Cheon, C.-Y., Lin, X. et al. Switching on and off the spin polarization of the conduction band in antiferromagnetic bilayer transistors. *Nat. Nanotechnol.* **20**, 609–616 (2025).

24. Pang, J., Niu, X., Zhang, M., Tang, Y., Zhang, Y. et al. Electric-field-induced formation and annihilation of skyrmions in a two-dimensional magnet. *Phys. Rev. B* **108**, 134430 (2023).

25. He, Z., Du, W., Dou, K., Dai, Y., Huang, B. et al. Ferroelectrically tunable magnetic skyrmions in two-dimensional multiferroics. *Mater. Horiz.* **10**, 3450–3457 (2023).

26. Cui, Q., Zhu, Y., Jiang, J., Liang, J., Yu, D. et al. Ferroelectrically controlled topological magnetic phase in a Janus-magnet-based multiferroic heterostructure. *Phys. Rev. Res.* **3**, 043011 (2021).

27. Dou, K., Du, W., Dai, Y., Huang, B. & Ma, Y. Two-dimensional magnetoelectric multiferroics in a MnSTe/$In_2Se_3$ heterobilayer with ferroelectrically controllable skyrmions. *Phys. Rev. B* **105**, 205427 (2022).

28. He, Z., Dou, K., Du, W., Dai, Y., Huang, B. et al. Multiple topological magnetism in van der Waals heterostructure of $MnTe_2$/$ZrS_2$. *Nano Lett.* **23**, 312–318 (2023).

29. Huang, K., Schwartz, E., Shao, D.-F., Kovalev, A. A. & Tsymbal, E. Y. Magnetic antiskyrmions in two-dimensional van der Waals magnets engineered by layer stacking. *Phys. Rev. B* **109**, 024426 (2024).

30. Wu, Y. et al. Voltage-controlled topological spin textures in the monolayer limit. *Nat. Commun.* **17**, 2923 (2026).

31. Sun, W., Wang, W., Li, H., Zhang, G., Chen, D. et al. Controlling bimerons as skyrmion analogues by ferroelectric polarization in 2D van der Waals multiferroic heterostructures. *Nat. Commun.* **11**, 5930 (2020).

32. Žutić, I., Matos-Abiague, A., Scharf, B., Dery, H. & Belashchenko, K. Proximitized materials. *Mater. Today* **22**, 85–107 (2019).

33. Lazić, P., Belashchenko, K. D. & Žutić, I. Effective gating and tunable magnetic proximity effects in two-dimensional heterostructures. *Phys. Rev. B* **93**, 241401 (2016).

34. Duan, C.-G., Velev, J. P., Sabirianov, R. F., Zhu, Z., Chu, J. et al. Surface magnetoelectric effect in ferromagnetic metal films. *Phys. Rev. Lett.* **101**, 137201 (2008).

35. Deng, Y., Yu, Y., Song, Y., Zhang, J., Wang, N. Z. et al. Gate-tunable room-temperature ferromagnetism in two-dimensional $Fe_3GeTe_2$. *Nature* **563**, 94–99 (2018).

36. Kim, J., Kim, K.-W., Kim, B., Kang, C.-J., Shin, D. et al. Exploitable magnetic anisotropy of the two-dimensional magnet $CrI_3$. *Nano Lett.* **20**, 929–935 (2020).

37. Euste, J. L., Hsouna, M. & Stojić, N. Transferable mechanism of perpendicular magnetic anisotropy switching by hole doping in $VX_2$ (X = Te, Se, S) monolayers. *Phys. Rev. B* **112**, 214423 (2025).

38. Guan, J., Chuang, H.-J., Zhou, Z. & Tománek, D. Optimizing charge injection across transition metal dichalcogenide heterojunctions: theory and experiment. *ACS Nano* **11**, 3904–3910 (2017).

39. Tenasini, G., Soler-Delgado, D., Wang, Z., Yao, F., Dumcenco, D. et al. Band gap opening in bilayer graphene–$CrCl_3$/$CrBr_3$/$CrI_3$ van der Waals interfaces. *Nano Lett.* **22**, 6760–6766 (2022).

40. Rizzo, D. J., Seewald, E., Zhao, F., Cox, J., Xie, K. et al. Engineering anisotropic electrodynamics at the graphene/CrSBr interface. *Nat. Commun.* **16**, 1853 (2025).

41. Kudrynskyi, Z. R., Bhuiyan, M. A., Makarovsky, O., Greener, J. D. G., Vdovin, E. E. et al. Giant quantum Hall plateau in graphene coupled to an InSe van der Waals crystal. *Phys. Rev. Lett.* **119**, 157701 (2017).

42. Singh, C. K. & Kabir, M. Long-range anisotropic Heisenberg ferromagnets and electrically tunable ordering. *Phys. Rev. B* **103**, 214411 (2021).

43. Verzhbitskiy, I. A. et al. Controlling the magnetic anisotropy in $Cr_2Ge_2Te_6$ by electrostatic gating. *Nat. Electron.* **3**, 460–465 (2020).

44. Sun, Q.-C. et al. Magnetic domains and domain wall pinning in atomically thin $CrBr_3$ revealed by nanoscale imaging. *Nat. Commun.* **12**, 1989 (2021).

45. Ghazaryan, D. et al. Magnon-assisted tunnelling in van der Waals heterostructures based on $CrBr_3$. *Nat. Electron.* **1**, 344–349 (2018).

46. Blöchl, P. E. Projector augmented-wave method. *Phys. Rev. B* **50**, 17953 (1994).

47. Kresse, G. & Hafner, J. Ab initio molecular-dynamics simulation of the liquid-metal-amorphous-semiconductor transition in germanium. *Phys. Rev. B* **49**, 14251 (1994).

48. Kresse, G. & Furthmüller, J. Efficient iterative schemes for ab initio total-energy calculations using a plane-wave basis set. *Phys. Rev. B* **54**, 11169 (1996).

49. Perdew, J. P., Burke, K. & Ernzerhof, M. Generalized gradient approximation made simple. *Phys. Rev. Lett.* **77**, 3865 (1996).

50. Grimme, S., Antony, J., Ehrlich, S. & Krieg, H. A consistent and accurate ab initio parametrization of density functional dispersion correction (DFT-D) for the 94 elements H–Pu. *J. Chem. Phys.* **132**, 154104 (2010).

51. Dudarev, S. L., Botton, G. A., Savrasov, S. Y., Humphreys, C. J. & Sutton, A. P. Electron-energy-loss spectra and the structural stability of nickel oxide: an LSDA+U study. *Phys. Rev. B* **57**, 1505–1509 (1998).

52. Gong, C., Kim, E. M., Wang, Y., Lee, G. & Zhang, X. Multiferroicity in atomic van der Waals heterostructures. *Nat. Commun.* **10**, 2657 (2019).

53. Evans, R. F. L., Fan, W. J., Chureemart, P., Ostler, T. A., Ellis, M. O. A. et al. Atomistic spin model simulations of magnetic nanomaterials. *J. Phys. Condens. Matter* **26**, 103202 (2014).

54. Berg, B. & Lüscher, M. Definition and statistical distributions of a topological number in the lattice O(3) σ-model. *Nucl. Phys. B* **190**, 412 (1981).